\documentstyle[pra,aps,draft]{revtex}
\begin{document}
\draft
\title{Tricritical Points in Random Combinatorics:
the $2+p$--SAT case}
\author{R\'emi Monasson \cite{rm} and Riccardo Zecchina \cite{rz}}
\address{ \cite{rm} Laboratoire de Physique Th\'eorique de l'ENS,
24 rue Lhomond, 75231 Paris cedex 05, France\\
\cite{rz} International Centre for Theoretical Physics, Strada
Costiera 11, P.O. Box 586, 34100 Trieste, Italy}

\maketitle
\begin{abstract}
The (2+p)-Satisfiability (SAT) problem interpolates between different
classes of complexity theory and is believed to be of basic interest
in understanding the onset of typical case complexity in random
combinatorics.  In this paper, a tricritical point in the phase
diagram of the random $2+p$-SAT problem is analytically computed using
the replica approach and found to lie in the range $2/5 \le p _0 \le
0.416$.  These bounds on $p_0$ are in agreement with previous numerical
simulations and rigorous results.
\end{abstract}


\section{Introduction}

The satisfiability (SAT) problem \cite{NPC} is the prototype of
NP--complete combinatorial decision problems arising in theoretical
computer science.  Such decision problems are, by definition, the most
difficult problems solvable in polynomial time by some ideal
non--deterministic algorithm \cite{NPC}.  In practice, however, real
algorithms may drastically change their performances depending on
whether the instances of the problem are highly constrained or not.
Therefore, the worst--case classification on which complexity theory
is founded does not necessarily capture the behaviour of search
algorithms in specific applications.
For example, random instances of NP--complete decision problems
undergo a dramatic change in the median time required for their
solution when the instances are generated at the boundary of a
critical region in the parameter space (for an 
introduction to these issues, see ref. \cite{hayes}).

A paradigm for such a behaviour is provided by the random 
K-Satisfiability (K-SAT) problem.  Briefly speaking, one is
given $N$ Boolean variables and a set of $M$ clauses to be satisfied
simultaneously. A clause refers to a logical constraint on $K$ Boolean
variables, randomly chosen among the $N$ ones. For large instances
($M,N \to \infty$), K-SAT exhibits a striking threshold phenomenon as
a function of the intensive ratio $\alpha =M/N$.  Numerical
simulations show that the probability of finding an assignment of the
Boolean variables satisfying all clauses, falls abruptly from one down
to zero when $\alpha$ crosses a critical value $\alpha_c (K)$ of the
number of clauses per variable \cite{nume}. This scenario is
rigorously established in the (Polynomial) $K=2$ case, where $\alpha
_c(2)=1$ \cite{goerdt}. For $K\ge 3$, much less is known; K($\ge
3$)-SAT belongs to the NP--complete class, roughly meaning that
running times of search algorithms are thought to scale exponentially
in $N$ when the problem instances are critically constrained. Recent
numerical works have provided an estimate for $\alpha _c (3) \simeq
4.2-4.3$ \cite{nume}.

A statistical mechanics approach has been attempted to get insights on
the K-SAT problem \cite{tish,MZI,MZII}. These studies relie on the
correspondence between solutions and ground--states of  diluted spin--glass
like cost--energy functions. Threshold phenomena therefore
correspond to zero temperature critical points in the phase diagram
of the associated spin glass model. Replica Symmetric (RS) theory
gives the correct value of the threshold for $K=2$ but fails in
predicting the critical $\alpha_c$ for $K \ge3$ \cite{MZI,MZII}. This
stems from the nature of the transition taking place at $\alpha_c$,
which is continuous for $K=2$ and appears discontinuous when $K\ge
3$. In the latter case, the precise location of the critical point for
the first order transition would require an appropriate replica
symmetry breaking scheme.  For interacting models with finite
connectivity, the latter issue is still an open problem under many
aspects \cite{eilat}.

Very recently \cite{MZKST}, it has been suggested that the particular
nature -- continuous or discontinuous -- of the phase transition
taking place at the threshold could be strictly
connected with the appearance of computationally hard instances, and
hence to the onset of exponential regimes in search algorithms
\cite{AI}. Recent numerical studies on the so-called $2+p$-SAT
problem \cite{MZKST}, that smoothly interpolates between 2-SAT ($p
=0$) and 3-SAT ($p =1$)\cite{MZII}, have strongly supported this
statement. It follows that the interest in the precise analytical
localisation of discontinuous transitions in random SAT models goes
much beyond the purely technical aspects of the replica formalism.

In this paper, we present the analytical calculation of the
tricritical point $p _0$ of the $2+p$-SAT model, separating
second-order phase transitions ($0\le p < p _0$) from first-order ones
($p _0 < p \le 1$). In section II, we recall the definition of the
$2+p$-SAT model.  The main steps of the statistical physics analysis
are exposed in Section III. In section IV, we study the critical
region and establish the self-consistent equations fulfilled by the
order parameter at threshold. We analyse these equations and show that
$2/5\le p_0 \le 0.416$. In conclusion, we underline the agreement between our
result and some recent mathematical study on the $2+p$-SAT model.

\section{Presentation of the $2+p$--SAT model}

The $2+p$-SAT model is a mixed version of 2-SAT and 3-SAT including
$(1-p)M$ (resp. $p M$) clauses constraining two (resp. three) Boolean
variables \cite{MZII}.

To start with, we consider a set of $N$ Boolean variables
$\{x_i=0,1\}_{i=1,\ldots,N}$. We first randomly choose $2$ among the
$N$ possible indices $i$ and then, for each of them, a literal $z_i$
that is the corresponding $x_i$ or its negation $\bar x_i$ with equal
probabilities one half. A clause $C$ is the logical OR of the $2$
previously chosen literals, that is $C$ will be true (or satisfied) if
and only if at least one literal is true.  Next, we repeat this
process to obtain $(1-p) M$ independently chosen clauses
$\{C_\ell\}_{\ell=1,\ldots,(1-p) M}$ and ask for all of them to be
true at the same time, i.e. we take the logical AND of the $M$ clauses
thus obtaining a Boolean expression in the so called Conjunctive
Normal Form (CNF).  The resulting 2--CNF formula $F_2$ may be written as
\begin{equation}
F_2 =\bigwedge_{\ell =1}^ {(1-p)M} C_\ell =\bigwedge_{\ell = 1}^
{(1-p)M}\;\;\left(
\bigvee_{i =1}^2 z_i ^{(\ell )}\right) \;\;\;,
\label{Fcnf2}
\end{equation}
where $\bigwedge$ and $\bigvee$ stand for the logical AND and OR
operations respectively. 

Then, using the above prescription, we generate a 3--CNF, hereafter
called $F_3$ including $p M$ clauses of length three.  The resulting
Boolean formula $F$ that we shall analyze, reads $ F=F_2 \wedge F_3$.
A logical assignment of the $\{x_i\}$'s satisfying all clauses, that
is evaluating $F$ to true, is called a solution of the
satisfiability problem. If no such assignment exists, $F$ is said
to be unsatisfiable.
It is worth noticing that as far as the complexity classification of the
problem is concerned, for any $p>0$ any instance of the model contains a
$3$--CNF sub--formula, therefore proving that the problem itself belongs to
the NP--complete class. 

This model has a threshold behaviour as usual K--SAT instances
\cite{MZKST,Optas} at a critical ratio $M/N = \alpha_c (p)$,
with $\alpha_c (0)=1$ and
$\alpha_c (1)= \alpha_c^{3sat} \simeq 4.2-4.3$. The critical ratio
is obviously bounded from above by $\alpha_c(p)\le 1/(1-p)$, 
obtained from the requirement that $F_2$ has to be almost surely
satisfiable. We shall show in the following that
\begin{equation}
\alpha_c (p) = \frac{1}{1-p} \qquad ,
\qquad ( 0\le p < p _0 ) ,
\label{satur}
\end{equation}
i.e. that the upper bound is reached when $p$ is smaller that
a value $p_0$ lying in the range 
\begin{equation}
0.4 \le p_0 \le 0.416 \qquad .
\end{equation}
Most remarkably, since an earlier presentation of our 
result\cite{MZKST}, a rigorous proof of the equality (\ref{satur}) 
has been derived for $p \le 2/5$ based on the analysis of the so-called
unit clause algorithm \cite{Optas}.

\section{Statistical mechanics analysis}

\subsection{The energy-cost function}

The above mixed random SAT problem can be mapped onto a diluted spin
cost--energy function upon introducing the spin variables,
$S_i=1$ if the Boolean variable $x_i$ is true, $S_i=-1$ if $x_i$ is
false, and by taking into account the clauses through an $M\times N$
random matrix $C$ where $C_{\ell ,i}=-1$
(respectively +1) if clause $C_l$ contains $\bar x_i$ (resp.\ $x_i$), 0
otherwise. It can be checked easily that $\sum _{i=1}^N C_{\ell i}
S_i$ equals the number of wrong literals in clause $\ell$.
Then the cost--energy function
\begin{equation}
E[C,S]=  \sum _{\ell =1}^{(1-p)M} \delta \left[ \sum _{i=1}^N C_{\ell i}
S_i ; -2 \right] + \sum _{\ell =(1-p)M+1}^{M} \delta \left[ 
\sum _{i=1}^N C_{\ell i} S_i ; -3 \right] \quad ,
\label{cost}
\end{equation}
where $\delta [.;.]$ denotes the Kronecker function,
counts the number of violated clauses in the CNF Boolean 
expression $F$ for logical assignment $S$.
The ground state (GS) energy of the cost function (\ref{cost}), i.e. its
minimum over $S$ at fixed $C$, encodes for the existence of satisfying
assignments (zero violated clauses, $E_{GS}=0$) or, if not, for the 
minimum number ($E_{GS}>0$) of violated clauses.

It is worth noticing that, in addition to usual two--spins interactions
that give rise to continuous phase transitions \cite{sk},
the energy (\ref{cost}) involves three-spins interactions due to the
presence of 3--clauses. The latter can generate discontinuous phase
transitions at sufficiently high concentration, i.e. for large enough 
$p$ \cite{Gardner}. The value of the tricritical point $p_0$ separating
the second order phase transitions from the first order ones on the
threshold line $\alpha _c(p)$ we want to calculate in the following.

\subsection{The average over the disorder}

Resorting to the replica method for diluted spin--glasses and
following ref.\cite{MZII},  one proceeds by computing the model 
``free--energy'' density at inverse temperature $\beta$, averaged over 
the clauses distribution $F(\beta)=-\frac{1}{\beta N}\overline{\ln Z[C]}$
where $Z[C]$ is the partition function.
The overline denotes the average over the random clauses matrices 
$C$ and is performed using the 
replica trick $\overline{\ln Z}=\lim_{n
\to 0} (\overline{Z^n}-1)/n$, starting from integer values of $n$. The
typical properties of the ground state, i.e. the internal energy and the
entropy, are recovered in the $\beta \to \infty$ limit.

To express the $n$-th moment of the partition function, it results
convenient to use the multi-level gas formalism proposed in
\cite{eilat}. The replicated theory is equivalent to a gas of $N$
particles occupying $2^n$ levels labelled by $n$-binary component 
vectors $\vec \sigma = (\sigma
_1 =\pm 1, \sigma _2 =\pm 1,\ldots , \sigma _n =\pm 1)$. 
Calling $\rho (\vec \sigma )$ the population, that is the fraction of
particles, on level $\vec \sigma $, the energy of the gas per particle
reads after some simple algebra exposed in Appendix,
\begin{eqnarray}
E_{gas} [\rho ] &=& - \frac{\alpha}{\beta}\; (1-p) \ \ln \left[ \sum
_{\vec \sigma , \vec \tau} \rho (\vec \sigma ) \rho (\vec \tau ) \
\exp \left( - \beta \sum _{a=1} ^n \delta [ \sigma _a ; 1] \delta [
\tau _a ; 1] \right) \right] \nonumber \\ && - \frac{\alpha}{\beta}
\;p \ \ln \left[ \sum _{\vec \sigma , \vec \tau , \vec \omega } \rho
(\vec \sigma ) \rho (\vec \tau ) \rho (\vec \omega ) \ \exp \left( -
\beta \sum _{a=1} ^n \delta [ \sigma _a ; 1] \delta [\tau _a ; 1]
\delta [\omega _a ; 1] \right) \right] \qquad ,
\label{expre}
\end{eqnarray}
with the symmetry constraint $\rho (\vec \sigma ) = \rho ( - \vec 
\sigma )$.
The stationary distribution $\rho _s$ of the level populations $\rho$ in the
thermodynamic limit $N\to \infty$ is obtained by balancing the
above energetic interactions and the
mixing entropy (per particle) \cite{eilat}
\begin{equation}
S_{gas} [\rho ] = - \sum _{\vec \sigma } \rho (\vec \sigma ) \ln
\rho (\vec \sigma ) \qquad ,
\end{equation}
that is minimising $E_{gas} [\rho ] -  S_{gas} [\rho ] /\beta$.  
The dominant contribution to $\overline{Z^n}$ is then given by
\begin{equation}
\overline{Z^n} \simeq \exp \left( - \beta N  \left[
E_{gas} [\rho _s] - \frac 1\beta S_{gas} [\rho _s] \right] \right)
\qquad .
\end{equation}
The determination of the saddle-point $\rho _s (\vec \sigma )$ is very
difficult in general but can be performed under some simplifying
assumptions. 

\subsection{The replica symmetric theory}

In the replica symmetric (RS) hypothesis, one looks for a stationary
distribution $\rho _s (\sigma _1 ,\sigma _2 ,\ldots ,\sigma _n )$
invariant under any permutation of the $n$ replicas. Therefore,
$\rho _s (\vec \sigma )$ depends on its argument through $\sum _{a=1}
^n \sigma _a$ only. This allows the introduction of a generating
function $R(z)$,
\begin{equation}
\rho _s (\sigma _1 ,\sigma _2 ,\ldots ,\sigma _n ) =
\int _{-\infty} ^\infty dz \; R(z) \prod _{a=1} ^n
\left( \frac{ e^{\beta z \sigma _a /2 }} { e^{\beta z /2}
+ e^{-\beta z /2} } \right) \qquad ,
\end{equation}
which becomes the Laplace transform of the populations $\rho _s$  in
the limit $n\to 0$ \cite{MZII}. Note that, since the sum of the
fractions $\rho$ equals one, $R(z)$ is normalized to unity.

The minimisation condition over $\rho _s$ yields a self-consistent
equation for the function $R(z)$. In the limits of interest $n\to 0$
and $\beta \to \infty$, this equation reads, see \cite{MZII},
\begin{eqnarray}
R(z) = && \int _{-\infty} ^\infty \frac{du}{2\pi } \cos (u z) \exp
\left\{ - \alpha (1-p) +2 \alpha (1-p) \int _0 ^\infty dz_1 R(z_1)
\cos ( u \min (1,z_1 )) \right. \nonumber \\
&& \left. - \frac 34 \alpha p + 3 \alpha p \int _0 ^\infty dz_1 dz_2
R(z_1) R(z_2 )\cos ( u \min (1,z_1 ,z_2 )) \right\} \qquad .
\label{eqr}
\end{eqnarray}
The interpretation of $R(z)$ is transparent within the cavity
approach~: it is the probability distribution of the effective fields
$z$ seen by the spins \cite{MZII}. In other words, $R(z)$  
accounts for  the histogram $P(\ll S \gg )$ of the thermal average 
values of the variables through the relation
$\ll S_i \gg = \tanh (\beta z_i /2)$.

\section{Analysis of the critical region}

\subsection{The order parameter at threshold}

When $\alpha < \alpha _c (p)$, that is for weakly constrained formulas, 
the stable solution of (\ref{eqr}) is $R(z)=\delta(z)$ since the
number of fully constrained spins in the ground state is not extensive
in $N$ \cite{MZII,MZKST}. Let us now fix $p$ to a small
value. According to the discussion of the previous Section, the SAT/UNSAT
transition is thought to be of the second order. We thus 
consider some small (and even) fluctuations $\mu (z) = R(z) - \delta (z) $ 
around the solution of the SAT phase. From (\ref{eqr}), we find
\begin{equation}
\mu (z) = \int_0 ^\infty dz_1 \; \tilde L(z , z_1 ) \mu (z_1) +
\int_0 ^\infty dz_1 dz_2 \; \tilde M(z,z_1 ,z_2 ) \mu (z_1) \mu (z_2)
+ O( \mu ^3) \quad ,
\end{equation}
for all $z$ where
\begin{eqnarray}
\tilde L(z , z_1 ) &=& \alpha (1-p) \sum _{\sigma _1 = \pm 1}
\delta ( z - \sigma _1 \min (1, z_1 )) \nonumber \\
\tilde M(z,z_1 ,z_2 ) &=& \frac 32 \alpha p \sum _{\sigma _1 = \pm 1}
\delta ( z - \sigma _1 \min (1, z_1,z_2 )) \nonumber \\
&& + \frac 12 \alpha ^2 (1-p)^2 \sum _{\sigma _1 , \sigma _2 = \pm 1}
\delta ( z - \sigma _1 \min (1, z_1)- \sigma _2 \min (1, z_2))
\qquad .
\label{expa}
\end{eqnarray}
Let us restrict to $z 
\in [0;1[$ \footnote{Equation (\ref{eqr}) is indeed a self-consistent
constraint on $R(z)$ on this range only, see \cite{MZII}.}.
The inspection of the linear term in (\ref{expa}) 
shows that the threshold is given by (\ref{satur}).  
Next, we expand around the latter by posing $\alpha=\alpha _c (p)+x$, 
$\mu (z) = x\; \eta (z) + O(x^2 )$ and obtain, when $x \to 0$, 
\begin{equation}
0 = (1-p) \; \eta (z) +
\int_0 ^\infty dz_1 dz_2 \; M(z,z_1 ,z_2 ) \; \eta (z_1) \eta (z_2)
 \quad , 
\label{eqeta}
\end{equation}
where the kernel of the quadratic form reads
\begin{eqnarray}
M(z,z_1 ,z_2 ) &=& \frac {3 p}{2(1-p)} \sum _{\sigma _1 = \pm 1}
\delta ( z - \sigma _1 \min (1, z_1,z_2 )) \nonumber \\
&& + \frac 12 \sum _{\sigma _1 , \sigma _2 = \pm 1}
\delta ( z - \sigma _1 \min (1, z_1)- \sigma _2 \min (1, z_2))
\qquad .
\label{expa1} 
\end{eqnarray}
Note that the positivity of the probability distribution $R$ imposes
$\eta (z) \ge 0$ for $z\ne 0$. Furthermore, the normalization of
$R$ implies that
\begin{equation}
\int _{-\infty} ^\infty dz \; \eta (z) = 0 \qquad .
\label{zero}
\end{equation}
Consequently, $\eta (z)$ includes a Dirac peak in $z=0$ with a
negative weight $- \eta _0 $, $\eta _0 \ge 0$.

\subsection{Discretisation of the self-consistent equations}

Within the iterative scheme for the RS solution discussed in \cite{MZII},
we can discretise the above equation and look for an exact solution
of the form 
\begin{equation}
\eta (z) = -\eta _0 \; \delta (z) + \sum _{\ell \ne 0 }
\eta _\ell \; \delta \left( z - \frac{\ell}{q} \right) \qquad ,
\label{ordre}
\end{equation}
In the above equation, $1/q$
is the resolution of the effective field which eventually goes to zero.
The self-consistent equations for the coefficients $\eta _\ell$'s,
$\ell =0 ,1, \ldots , q-1$ are
easily obtained from (\ref{eqeta}),
\begin{equation}
(1-p) \eta _0 = \frac 34\; \frac{1-2p}{1-p} \eta _0 ^2 
- \eta _0 \sum _{j=1}^{q-1} \eta _j + \sum _{j=1}^{q-1} \eta _j ^2 + \left( 
\sum _{j=1}^{q-1} \eta _j  \right) ^2
\label{t1}
\end{equation}
and, for $\ell=1,\ldots , q-1$,
\begin{eqnarray}
(1-p) \eta _\ell &=& \eta _\ell \left\{ \eta _0 +  \frac 32
 \frac{p}{1-p} \left[ -\eta _0 +  2 \sum
_{j=1}^{\ell -1} \eta _j + \eta _\ell \right] \right\}
 \nonumber \\ 
&& - \frac 12 \sum _{j=1} ^{\ell -1} \eta _j \eta _{\ell - j} -
\sum _{j=1} ^{q -\ell -1} \eta _j \eta _{\ell+ j}
+ \eta _{q-\ell} \left( \sum _{j=1}^{q-1} \eta _j  - 
\frac 12 \eta _0 \right) \qquad .
\label{t2}
\end{eqnarray}

\subsection{Homogeneous equations at tricriticallity}

The onset of first order transition corresponds to the smallest value
of $p$ for which $\eta (z)$ diverges. Let us call $p_0 (q)$ the tricritical
point for a resolution of the field $1/q$. 
When $q=1$, equation (\ref{t1}) gives
\begin{equation}
\eta _0 = \frac {4(1-p)^2}{3(1-2p)} \qquad,
\label{sol1}
\end{equation}
leading to $p_0(1)=1/2$. When increasing $q$, one gets smaller and
smaller values for $p_0(q)$, e.g. $p_0 (2) = 0.4614 , p_0 (3)= 0.4484,
\ldots$. When approaching $p_0(q)$ from below, the weights of the Dirac peaks 
always diverge according to
\begin{equation}
\eta _\ell (p) \simeq \frac{\Omega _\ell}{ p_0 (q) - p }\ , \qquad 
p \to p_0 (q) ^-  \qquad ,
\label{scal}
\end{equation}
as can be explicitly checked on (\ref{sol1}) for $q=1$ and $\ell=0$. 
Therefore, the amplitudes $\Omega _\ell $ have to 
satisfy the {\em homogeneous}
versions of equations (\ref{t1}, \ref{t2}),
\begin{equation}
0 = \frac 34\; \frac{1-2p}{1-p} \Omega _0 ^2 - \Omega _0 \sum
_{j=1}^{q-1} \Omega _j + \sum _{j=1}^{q-1} \Omega _j ^2 + \left( \sum
_{j=1}^{q-1} \Omega _j \right) ^2
\label{ome1}
\end{equation}
and, for $\ell=1,\ldots , q-1$,
\begin{eqnarray}
0 &=& \Omega _\ell \left\{ \Omega _0 + \frac 32 \frac{p}{1-p}
 \left[ -\Omega _0 + 2 \sum _{j=1}^{\ell -1} \Omega _j + \Omega _\ell
 \right] \right\} \nonumber \\ && - \frac 12 \sum _{j=1} ^{\ell -1}
 \Omega _j \Omega _{\ell - j} - \sum _{j=1} ^{q -\ell -1} \Omega _j
 \Omega _{\ell+ j} + \Omega _{q-\ell} \left( \sum _{j=1}^{q-1} \Omega
 _j - \frac 12 \Omega _0 \right) \qquad .
\label{ome2}
\end{eqnarray}
The tricritical point $p_0$ is the smallest value of $p$ for which the
quadratic forms in (\ref{ome1},\ref{ome2}) have a non zero solution
$\Omega _\ell$. In the above equations, we
can choose $\Omega _0 =1$ arbitrarily and we are left with $q$ coupled
equations for $p_0$ and the $q-1$ amplitudes $\Omega _\ell$, 
$\ell =1,\ldots ,q-1$.

\subsection{Lower bound to the tricritical point}

We now focus upon the self-consistent equation
(\ref{ome1}) that we rewrite as follows,
\begin{equation}
\frac{5 p -2}{4 (1-p)}\; \Omega _0 ^2 = 
\left( \frac 12 \Omega _0 - \sum _{j=1}^{q-1} \Omega
 _j \right) ^2 + \sum _{j=1}^{q-1} \Omega
 _j ^2
\qquad ,
\label{eta0}
\end{equation}
from which the lower bound $2/5 \le p_0$ is immediately derived.
Furthermore, this 
lower bound can be reached if and only if $\eta (z)$ at the tricritical point
vanishes outside of the interval $]-1;1[$ and no Dirac distribution are 
present in the continous limit $q\to \infty$.

\subsection{Upper bound to the tricritical point}

If $\{\tilde 
\Omega _\ell\} $ is a solution 
of equations (\ref{ome1}, \ref{ome2}) for a given pair of parameters
$(p,q=\tilde q)$, so is
$\{\Omega _{ 2 \ell } = \tilde \Omega _\ell ,
\Omega _{ 2 \ell + 1 } = 0 \} $ for $(p , q = 2 \tilde q )$.
Thus, $p_0 (q) \ge p_0 ( 2 q)\ge \ldots
 \ge p_0 $ for any finite $q$, defining 
a sequence of more and more refined upper bounds to $p_0$.
We have then obtained the numerical values of $p_0(q)$
for $q=1, \ldots ,120$. It appears that $p_0 (q)$ indeed 
decreases with $q$ and equals 0.4158 for $q=120$, 
giving a numerical upper bound to $p_0$.

The convergence of $p_0 (q)$ down to its limit value $p_0$ is very
slow and seems to display some power law effects, see figure 1. At
first sight, the numerical prediction for $p_0$ is close to $0.41$, a
value close but higher than the lower bound $2/5$.

\section{Conclusion}

To end with, few observations are in order.
The above results have been derived within an iterative RS scheme allowing 
for more and more refined effective field resolutions.
With the simplest choice of integer fields, the value of $p_0$ would have 
been $\frac{1}{2}$, a wrong result which tells us that there must exist 
other non--integer contibutions to $R(z)$.
The appearance of non integer effective fields has recently been shown 
to reflect the existence of Replica Symmetry Breaking (RSB).
Further work will be neccessary to elucidate the role of RSB effects on 
the structure of the solutions (in principle, even the calculation of $p_0$ 
could be affected).
The rigorous results discussed in ref. \cite{Optas}, show that the 
RS solution is exact at least up to $p<2/5$. Such probabilistic
results are based on the convergency analysis of a simple algorithm which
proceeds by successive simplifications of the Boolean formula originated by
fixing at random one variable at a time.
In ref. \cite{Optas} it is shown that for $\alpha < \alpha_c(p)$ and $p<2/5$,
the above algorithm has a finite probability of finding a satisfying 
assignment and hence the starting formula has to be satisfiable with 
probability one in the limit $N \to \infty$.
For $p<2/5$ the 3--clauses are ineffective even for a rather trivial
``dynamical process'' like the mentioned algorithm.
Such a result is indeed consistent with the idea that the nature of the phase
transition taking place at $\alpha_c(p)$ does not change at least up to
$p<2/5$. In the case $p_0>2/5$, as suggested by the RS solution,
it should be of interest to understand how one should modify the algorithm in
order to recover the statistical mechanics result.

Let us conclude by noticing that from a physical point of view, the nature of
the transition manifests itself through the appearance of a finite fraction
of completely constrained variables when crossing the threshold 
\cite{MZII,Antonio}.
Above $p _0$, this fraction discontinuously blows up at $\alpha _c $. The
narrow correspondence between this fact and the onset of computational
complexity shown by simulations \cite{MZKST} suggests that the underlying
mechanisms causing the increase of the typical computational search cost
could be related to the fact that search algorithms have to find the precise
values of a $O(N)$ number of Boolean variables through an extensive
enumeration.

{\bf Acknowledgments}~: We thank O. Dubois, S. Kirkpatrick, B. Selman for
valuable comments.
We also thank the Microsoft Theory Group for discussions and hospitality.
\appendix

\section{Calculation of the effective gas energy}

Consider $n$ Boolean assignments $S^a$, where $a=1,\ldots ,n$, each comprised
of $N$ binary spins. 
The replica method requires the computation of the average product of their Gibbs weights corresponding to energy (\ref{cost}),
\begin{equation}
z [S^a  ] = \overline{ \exp \left( -\beta \sum _{a=1} ^n
E[ C , S^a ] \right) } 
\end{equation}
factorises over the sets of two- and three-clauses due to the absence of any 
correlation in their probability distribution. Thus,
\begin{equation}
z [ S^a ] = \left( \zeta _2 [ S^a ] \right) ^{(1-p) M}
\    \left( \zeta _3 [ S^a ] \right) ^{p M}
\qquad .
\end{equation}
The single clause factors in the above formula are defined by (for $K=2,3$)
\begin{equation}
\zeta _K [ S^a ] = \overline{ \exp \left( -\beta \sum _{a=1} ^n
\delta \left[ \sum _{i=1}^N C_i S_i^a ; -K \right] \right) }
\qquad , \label{appe}
\end{equation}
where the bar denotes the unbiased average over the set of $2^K {N
\choose K}$
vectors of $N$ components $C_i =0 ,\pm 1$ and of squared norm equal to $K$.
Using the identity,
\begin{equation}
\delta \left[ \sum _{i=1}^N C_i S_i^a ; -K \right] =
\prod _{i / C_i \ne 0} \delta \left[ S_i^a ; -C _i \right] \qquad,
\end{equation}
we carry out the average over in disorder in (\ref{appe}) to obtain
\begin{equation}
\zeta _K [ S^a ] = \frac 1{2^K} \sum _{C_1 ,\ldots , C_K =\pm 1} \frac 1{N^K}
\sum _{i_1 ,\ldots , i_K =1} ^N \exp \left\{ -\beta \sum _{a=1} ^n
\prod _{\ell =1}^K \delta \left[ S_{i_\ell} ^a ; -C _\ell\right] \right\}
\quad ,
\end{equation}
to the largest order in $N$. Defining $\rho (\vec \sigma )$ as the fraction of
spins $( S^1_i ,\ldots ,S^n _i )$ equal to $(\sigma ^1 ,\ldots ,\sigma ^n )$
\cite{eilat}, we rewrite $\zeta _K [ S^a ] =\zeta _K [ \rho ]$ with
\begin{equation}
\zeta _K [ \rho ] = \frac 1{2^K} \sum _{C_1 ,\ldots , C_K =\pm 1} 
\sum _{\vec \sigma _1 , \ldots , \vec \sigma _K}  
\rho (- C_1 \;\vec \sigma _1  ) \ldots 
\rho (- C_K\; \vec \sigma _K )\exp \left\{ 
-\beta \sum _{a=1} ^n \prod _{\ell =1}^K \delta \left[ \sigma _\ell
 ^a ; 1 \right] \right\}
\quad .
\end{equation}
Notice that $\rho (\vec \sigma ) = \rho (-\vec \sigma ) $ due to the
even distribution of the disorder $C$. The final expression of the effective
gas energy per particle, defined as $E _{gas} [\rho ] = - \log
z [ S^a ]  / \beta N $ is given in (\ref{expre}).

\begin{figure}[tbp]
\caption{Plot of $p_0(q)$  versus $q$.
The dashed line is the lower bound $p_0 =2/5$.
}
\label{fig1}
\end{figure}

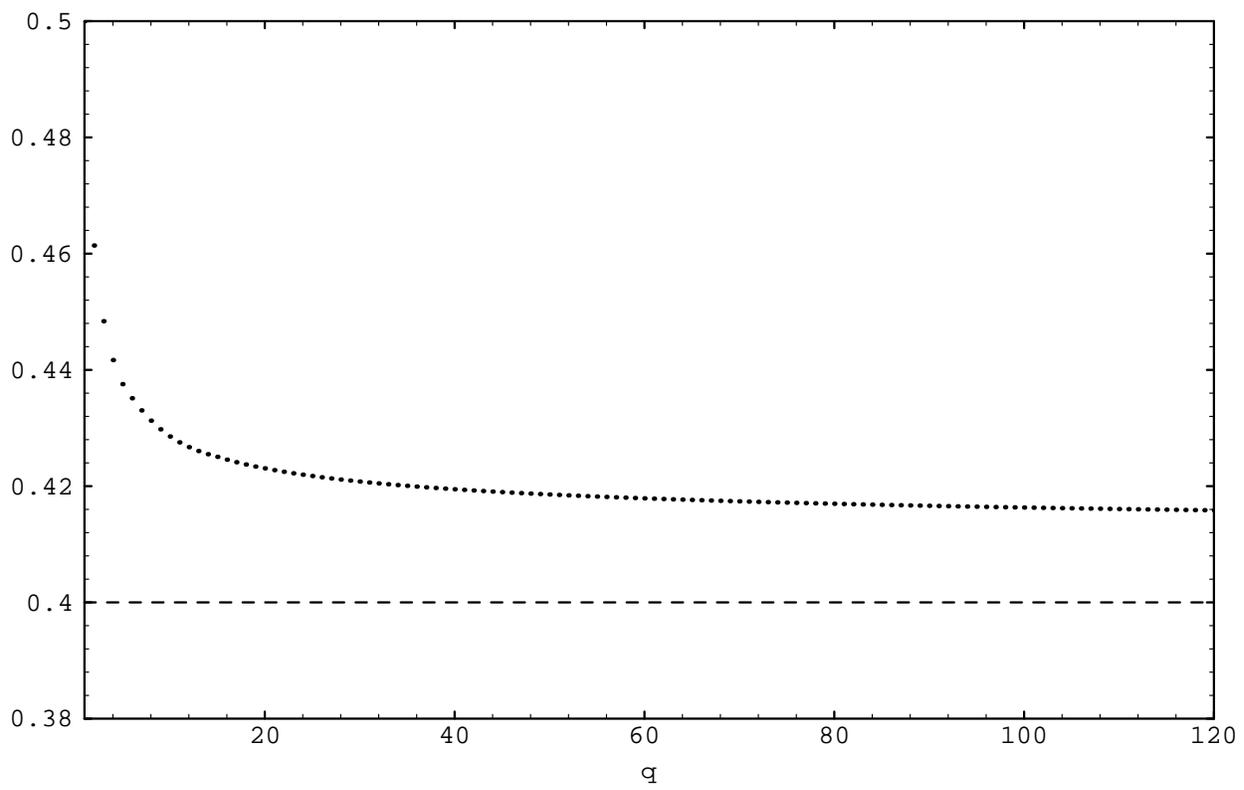
\end{document}